\documentclass[10pt,conference]{IEEEtran}\IEEEoverridecommandlockouts
\usepackage{epsfig,graphicx,subfigure,psfrag,amsmath,cases,bm}
\usepackage{latexsym,amssymb,amsmath,epsfig,subfigure,algorithm,mathtools}
\usepackage{algorithmic}
\usepackage{color}
\usepackage{url}
\usepackage{scrtime}

\author{\IEEEauthorblockN {Yan Sun\IEEEauthorrefmark {1}, Chao Wang\IEEEauthorrefmark {1}, Huan Cai\IEEEauthorrefmark {2}, Chunming Zhao\IEEEauthorrefmark {2}, Yiqun Wu\IEEEauthorrefmark {1}, Yan Chen\IEEEauthorrefmark {1}}
	
	\IEEEauthorrefmark {1}Huawei Technology \hspace*{30mm}
	\IEEEauthorrefmark {2}Southeast University, China
}

\title{Deep Learning Based Equalizer for MIMO-OFDM Systems with Insufficient Cyclic Prefix }

\newtheorem{T-Prob}{Transformed Problem}

\DeclareMathOperator{\diag}{\mathrm{diag}}
\DeclareMathOperator{\Diag}{\mathrm{Diag}}
\DeclareMathOperator{\FFT}{\mathrm{FFT}}

\newcommand{\abs}[1]{\lvert#1\rvert}
\newcommand{\norm}[1]{\lVert#1\rVert}
\begin{document}
	
	\maketitle
	
	\begin{abstract}
		In this paper, we study the equalization design for multiple-input multiple-output (MIMO) orthogonal frequency division multiplexing (OFDM) systems with insufficient cyclic prefix (CP). 
		In particular, the signal detection performance is severely impaired by inter-carrier interference (ICI) and inter-symbol interference (ISI) when the multipath delay spread exceeding the length of CP. 
		To tackle this problem, a deep learning-based equalizer is proposed for approximating the maximum likelihood detection. Inspired by the dependency between the adjacent subcarriers, a computationally efficient joint detection scheme is developed. Employing the proposed equalizer, an iterative receiver is also constructed and the detection performance is evaluated through simulations over measured multipath channels. Our results reveal that the proposed receiver can achieve significant performance improvement compared to two traditional baseline schemes. 
	\end{abstract}

	\section{Introduction}
	Orthogonal frequency division multiplexing (OFDM) is the fundamental transmission technique for the long term evolution (LTE) and the 5th generation new radio (5G-NR) systems, due to the great improvement in spectral efficiency and the flexibility they offer for resource allocation \cite{sun2016optimalJournal}.
	The promised gain of OFDM is realized by adding cyclic prefix (CP) at the beginning of each OFDM symbol, in order to guarantee the orthogonality among subcarriers \cite{nee2000ofdm}. In particular, the length of CP is expected to be greater than the length of channel impulse response (CIR).
	However, in practice, the length of CP is fixed while the propagation environments are various where the multipath delay spread may exceed the length of CP \cite{zhu2000channel}. 
	As a consequence, the orthogonality among subcarrier are defeated and the resulting inter-carrier interference (ICI) and inter-symbol interference (ISI) will severely degrade the system performance. 
	Moreover, the interference becomes more severe in multiple-input multiple-output (MIMO) OFDM systems since multiple data streams are transmitted simultaneously which leads to the additional inter-antenna interference (IAI). 
	
	To overcome these issues, equalization designs for eliminating the ICI and the ISI caused by insufficient CP have been investigated in the literature \cite{zhu2000channel}\nocite{parsace2004mmse,molisch2007iterative}--\cite{nisar2007channel}. 
	The mathematical signal model of insufficient CP in single-input single-output (SISO) OFDM systems was first investigated in \cite{zhu2000channel}. 
	In \cite{parsace2004mmse},  a time-domain decision feedback equalizer (DFE) based on minimum mean square error (MMSE) criterion was proposed to decrease the power of ISI and ICI, whereas the performance is restricted by the feedback error. 
	In \cite{molisch2007iterative}, the authors proposed an equalization design based on iterative ICI cancellation, yet a considerable number of iterations are needed for convergence.
	The authors of \cite{nisar2007channel} studied the joint channel estimation and equalization design for the suppression of insufficient CP generated interference, illustrating the impact of channel estimation quality on detection performance. 
	Besides, the equalization design for handling insufficient CP issues for MIMO-OFDM was investigated in \cite{pham2016channel,jin2012interference}. 
	In \cite{pham2016channel}, an time-domain iterative equalization scheme was proposed for MIMO detection with insufficient CP which entailed a high computational complexity.
	In \cite{jin2012interference}, the transmit precoding was designed for nulling the ICI and the ISI. However, it requires larger number of antennas on the transmit side than the transmit side which is only applicable for the downlink transmission and cannot fully exploit the spatial diversity.
	Moreover, the above works in \cite{zhu2000channel}\nocite{parsace2004mmse,molisch2007iterative,nisar2007channel,pham2016channel}--\cite{jin2012interference} are designed based on conventional signal processing methods which entails high computational complexity when OFDM systems employ a large number of subcarriers.
	
	Recently, deep learning is emerging as a promising approach for communication signal processing owing to its strong capability in non-linear model approximation,  feature extraction, and optimal decision \cite{gunduz2019machine}. For instance,  deep learning-based schemes have been investigated for MIMO detection \cite{samuel2019learning}, channel coding \cite{kurka2020deepjscc}, and channel estimation \cite{yang2020graph}.
	Using  deep learning for handling interference caused by insufficient CP was first investigated in \cite{Zhang19AICPfree}.
	In particular, the authors of \cite{Zhang19AICPfree} proposed a CP-free OFDM system where a model-driven neural network inspired by orthogonal approximate message passing (OAMP) was designed to reduce the inherent ICI and ISI and to recover signals. 
	However, the work in \cite{Zhang19AICPfree} focuses on SISO-OFDM and cannot be applied in MIMO-OFDM systems. Besides, the proposed model-driven deep neural network (DNN) in \cite{Zhang19AICPfree} is based on OAMP algorithm which is not the optimal detection scheme in general. 
	Moreover, channel coding was not taken into account for performance evaluation in \cite{Zhang19AICPfree}, and hence the link-level performance is still not clear.
	
	In this paper, we address the above issues. To this end, we investigate the deep learning-based equalization algorithm design for MIMO-OFDM systems with insufficient CP. The proposed equalizer mimics the procedure of solving a maximum likelihood estimation problem. The dependency among the adjacent subcarriers is explored for reducing the computational complexity.
	The performance of the proposed equalizer is evaluated over the measured channel model with a long multipath delay spread.  
	

	\section{System Model}
	In this section, we present the considered MIMO-OFDM system model taking into account insufficient CP.
	In particular, we consider an uplink multiuser MIMO-OFDM system which consists of a BS and $U$ UL users. The BS is equipped with $N_\mathrm{R} > 1$ antennas. Each UL user is equipped with $N_\mathrm{t} \ge 1$ antennas and the total number of transmit antennas of all users is $N_\mathrm{T} = U N_\mathrm{t}$. The entire frequency band is partitioned into $N$ orthogonal subcarriers. 
	
	\subsection{Channel Model in Frequency Domain}
	We start the channel modeling from the single-input and single-output (SISO) case. 
	Data bits are mapped into complex constellations from the M-quadrature amplitude modulation (M-QAM) constellation set. Let $\dot{\mathbf{X}}_k = [\dot{X}_{k,1}, \cdots, \dot{X}_{k,N}]^T$ denotes the transmitted symbol sequence at the $k$-th OFDM symbols in frequency, where $\dot{X}_{k,i}, \forall i \in \{1,\ldots, N\}$ denotes the constellation value on subcarrier $i$. By $N$-point inverse fast Fourier transform (iFFT) and adding CP with length of $L_{\mathrm{CP}}$, the OFDM symbol $k$ in time-domain is transmitted over a channel $\mathbf{h}_k$, which is given by:
	\begin{eqnarray}
	\mathbf{h}_k = [h_{k}^1, h_{k}^2, \ldots, h_{k}^{L_\mathrm{h}}]^T,
	\end{eqnarray}
	where $L_{\mathrm{h}}$ denotes the length of the CIR of $\mathbf{h}_k$.
	At the BS side, the received signal is handled with removing CP and performing $N$-point FFT. 
	According to \cite{Cisek18Frequency}, the received $k$-th OFDM symbol for SISO system in frequency domain can be written as:
	\begin{eqnarray}\label{SISO-model}
	\dot{\mathbf{Y}}_k = \dot{\mathbf{H}}_k \dot{\mathbf{X}}_k - \underbrace{\dot{\mathbf{\Phi}}_k \mathbf{G} \dot{\mathbf{X}}_k}_{\mathrm{ICI}} + \underbrace{\dot{\mathbf{\Phi}}_{k-1} \dot{\mathbf{X}}_{k-1}}_{\mathrm{ISI}} + \dot{\mathbf{N}}_k
	\end{eqnarray}
	where 
	$\dot{\mathbf{Y}}_k = [Y_{k,1}, \cdots, Y_{k,N}]^T$ and $\dot{\mathbf{N}}_k$ denotes the additive noise in frequency domain. The channel matrix $\dot{\mathbf{H}}_k \in \mathbb{C}^{N \times N}$  is a diagonal matrix where its dominant elements are the FFT transform of $\mathbf{h}_k$, which is given by:
	\begin{eqnarray}\label{H-SISO}
	\dot{\mathbf{H}}_k = \diag(\FFT(\mathbf{h}_k)).
	\end{eqnarray}
	We define $\mathbf{F}$ as a unitary $N\times N$ FFT matrix with elements given by $[\mathbf{F}]_{n,m} = W^{nm}$, where $W=e^{-j\frac{2pi}{N}}$.
	Then, $\mathbf{G}$ in \eqref{SISO-model} is a diagonal matrix where diagonal elements are composed by the $L_{\mathrm{CP}}$-th column of the FFT matrix $\mathbf{F}$, i.e., $\mathbf{G} = \diag([W^0, W^{L_{\mathrm{CP}}},\cdots, W^{(N-1)L_{\mathrm{CP}}}]^T)$. 
	$\dot{\mathbf{\Phi}}_k^{(n,m)} \in \mathbb{C}^{N \times N}$ denotes the interference channel matrix between transmit antenna $m$ and receiving antenna $n$ for symbol $k$, and is written as 
	\begin{eqnarray} \label{Phi-SISO}
	\dot{\mathbf{\Phi}}_k = \mathbf{F}
	\begin{bmatrix}
	\mathbf{0}& 
	\begin{bmatrix}
	h_{k}^{L_{\mathrm{h}}} & h_{k}^{L_{\mathrm{h}}-1} & \cdots & h_{k}^{L_{\mathrm{CP}}+1}\\
	0 & h_{k}^{L_{\mathrm{h}}} & \cdots & h_{k}^{L_{\mathrm{CP}}+2} \\
	\vdots & 0 & \ddots & \vdots\\
	0 & 0 & \cdots & h_{k}^{L_{\mathrm{h}}} \\
	\end{bmatrix}\\
	\mathbf{0} & \mathbf{0}
	\end{bmatrix} \mathbf{F}^{H}.
	\end{eqnarray}
	
	Now, we extend the signal model for SISO-OFDM in \eqref{SISO-model} to MIMO-OFDM systems. In particular, the received $k$-th OFDM symbol at the BS can be written as:
	\begin{eqnarray} \label{Hk-MIMO}
	\mathbf{Y}_k = \mathbf{H}_k\mathbf{X}_k - \mathbf{\Phi}_k \mathbf{B} \mathbf{X}_k + \mathbf{\Phi}_{k-1}\mathbf{X}_{k-1} + \mathbf{N}_k
	\end{eqnarray}
	where 
	\begin{eqnarray}
	\mathbf{Y}_k = [Y_{k,1}^{(1)}, \ldots,Y_{k,N}^{(1)}, \ldots, Y_{k,1}^{(N_\mathrm{R})},\ldots,Y_{k,N}^{(N_\mathrm{R})}]^T,
	\end{eqnarray}
	$Y_{k,i}^{(n)}$ denotes the constellation value of the $n$-th antenna at the $k$-th received symbol on subcarrier $i$, and 
	\begin{eqnarray}
	\mathbf{X}_k = [X_{k,1}^{(1)}, \ldots,X_{k,N}^{(1)}, \ldots, X_{k,1}^{(N_\mathrm{T})},\ldots,X_{k,N}^{(N_\mathrm{T})}]^T,
	\end{eqnarray}
	$X_{k,i}^{(m)}$ denotes the complex value of the $m$-th transmit antenna at the $k$-th symbol on subcarrier $i$, respectively. 
	The channel matrix $\mathbf{H}_k \in \mathbb{C}^{N N_{\mathrm{R}}\times N N_{\mathrm{T}}}$ is written as
	\begin{eqnarray}
	\mathbf{H}_k &=& 
	\begin{bmatrix}
	\mathbf{H}_k^{(1,1)} & \cdots & \mathbf{H}_k^{(1,\mathrm{N}_T)}\\
	\mathbf{H}_k^{(2,1)} & \cdots & \mathbf{H}_k^{(2,\mathrm{N}_T)}\\
	\vdots & \cdots & \vdots \\
	\mathbf{H}_k^{(\mathrm{N}_R,1)} & \cdots & \mathbf{H}_k^{(\mathrm{N}_R,\mathrm{N}_T)}
	\end{bmatrix},
	\end{eqnarray}
	where $\mathbf{H}_k^{(n,m)}$ denotes the channel matrix for OFDM symbol $k$ from transmit antenna $m$ to receive antenna $n$ and can be expressed as the form of \eqref{H-SISO}.
	The interference channel matrix $\mathbf{\Phi}_k \in \mathbb{C}^{N N_{\mathrm{R}}\times N N_{\mathrm{T}}}$ is expressed as
	\begin{eqnarray}
	\mathbf{\Phi}_k &=& 
	\begin{bmatrix}
	\mathbf{\Phi}_k^{(1,1)} & \cdots & \mathbf{\Phi}_k^{(1,\mathrm{N}_T)}\\
	\mathbf{\Phi}_k^{(2,1)} & \cdots & \mathbf{\Phi}_k^{(2,\mathrm{N}_T)}\\
	\vdots & \cdots & \vdots \\
	\mathbf{\Phi}_k^{(\mathrm{N}_R,1)} & \cdots & \mathbf{\Phi}_k^{(\mathrm{N}_R,\mathrm{N}_T)}
	\end{bmatrix}
	\end{eqnarray}
	where $\mathbf{\Phi}_k^{(n,m)}$ denotes the interference channel matrix for OFDM symbol $k$ from transmit antenna $m$ to receive antenna $n$ and can be presented in the form of \eqref{Phi-SISO}. Besides, $\mathbf{B} = \Diag(\mathbf{G})$. In the next section, we study the equalization design using the signal model derived in 
	\eqref{Hk-MIMO}.
	
	\section{Iterative Receiver with DNN based Equalizer}
	In this section, we first formulate the equalization design as an optimization problem based on maximum-likelihood (ML) estimation criterion. Then, a model-driven DNN is proposed for imitating the procedure of solving the formulated ML detection problem, which approximates the optimal solution of the ML detection. In addition, the computational complexity of the proposed DNN is reduced via partial ISI and ICI cancellation. Moreover, the complete receiver structure is also presented.
	
	\subsection{Maximum-likelihood Equalization}
	Since the ISI on the intended symbol $k$ is mainly caused by the previous symbol $k-1$, we focus on the received symbol $k$ and $k-1$ in frequency domain which are given by:
	\begin{eqnarray}
	\label{Yk-1}
	\mathbf{Y}_{k-1} \hspace*{-2mm} & = & \hspace*{-2mm} \mathbf{H}_{k-1}\mathbf{X}_{k-1} - \mathbf{\Phi}_{k-1} \mathbf{B} \mathbf{X}_{k-1} + \mathbf{\Phi}_{k-2}\overline{\mathbf{X}}_{k-2}, \\
	\label{Yk}
	\mathbf{Y}_k \hspace*{-2mm} & = & \hspace*{-2mm} \mathbf{H}_k\mathbf{X}_k - \mathbf{\Phi}_k \mathbf{B} \mathbf{X}_k + \mathbf{\Phi}_{k-1}\mathbf{X}_{k-1}, 
	\end{eqnarray}
	where $\overline{\mathbf{X}}_{k-2}$ denotes the estimation value after pre-detection, for example, using conventional MMSE equalization or zero-forcing (ZF) equalization. The noise term $\mathbf{N}_k$ and $\mathbf{N}_{k-1}$ are omitted without affecting the equalization design. Then, equations \eqref{Yk-1} and \eqref{Yk} can rewritten as:
	\begin{eqnarray}
	\mathbf{R}_k = \mathbf{A}_k \mathbf{Z}_k,
	\end{eqnarray}
	where 
	\begin{eqnarray}
	\mathbf{R}_k & = & 
	\begin{bmatrix}
	\mathbf{Y}_{k-1} - \mathbf{\Phi}_{k-2} \overline{\mathbf{X}}_{k-2} \\
	\mathbf{Y}_k
	\end{bmatrix},\\
	\label{Ak}
	\mathbf{A}_k & = & 
	\begin{bmatrix}
	\mathbf{H}_{k-1} - \mathbf{\Phi}_{k-1} \mathbf{B} & \mathbf{0}\\
	\mathbf{\Phi}_{k-1} & \mathbf{H}_k - \mathbf{\Phi}_k \mathbf{B}
	\end{bmatrix},\\
	\mathbf{Z}_k & = &
	\begin{bmatrix}
	\mathbf{X}_{k-1} \\
	\mathbf{X}_{k}
	\end{bmatrix}.
	\end{eqnarray}
	Hence, the ML detection can be obtained by solving the following problem \cite{cho2010mimo}: 
	\begin{eqnarray}
	\label{origML}
	\hat{\mathbf{Z}}_k^{\mathrm{ML}} = \underset{\mathbf{Z}_k \in \mathcal{Z}_{\mathrm{QAM}}}{\arg \min} \,\, \norm{\mathbf{R}_k - \mathbf{A}_k \mathbf{Z}_k}^2
	\end{eqnarray}
	where $\hat{\mathbf{Z}}_k^{\mathrm{ML}}$ denotes the estimation in terms of ML criterion, and $\mathcal{Z}_{\mathrm{QAM}}$ denotes the QAM constellation set.
	It is known that the ML estimation problem in \eqref{origML} yields the optimal detection if transmit symbols follow uniform prior distribution which normally holds in practical communication systems \cite{cho2010mimo}.
	However, problem \eqref{origML} is a combinatorial optimization problem which entails an exhaustive search for finding the optimal solution.
	Moreover, the computational complexity of solving \eqref{origML} scales with the number of subcarriers $N$ which is an obstacle for the application in practice. Thus, in the following subsection, we propose a low-complexity DNN-based equalizer for approaching the optimal solution of the ML detection problem in  \eqref{origML}.
	
	\subsection{DNN-based Equalizer}
	In fact, the ICI and ISI is generated when passing OFDM symbols through multipath channels in time domain. After the FFT operation, the ICI and ISI impairs all subcarriers at the receiver side. Intuitively, joint estimation over all subcarriers yields the optimal detection performance. However, this leads to a very high computational complexity when the number of subcarriers $N$ is large.
	
	\begin{figure}
		\centering\vspace*{-2mm}
		\includegraphics[width=3.2in]{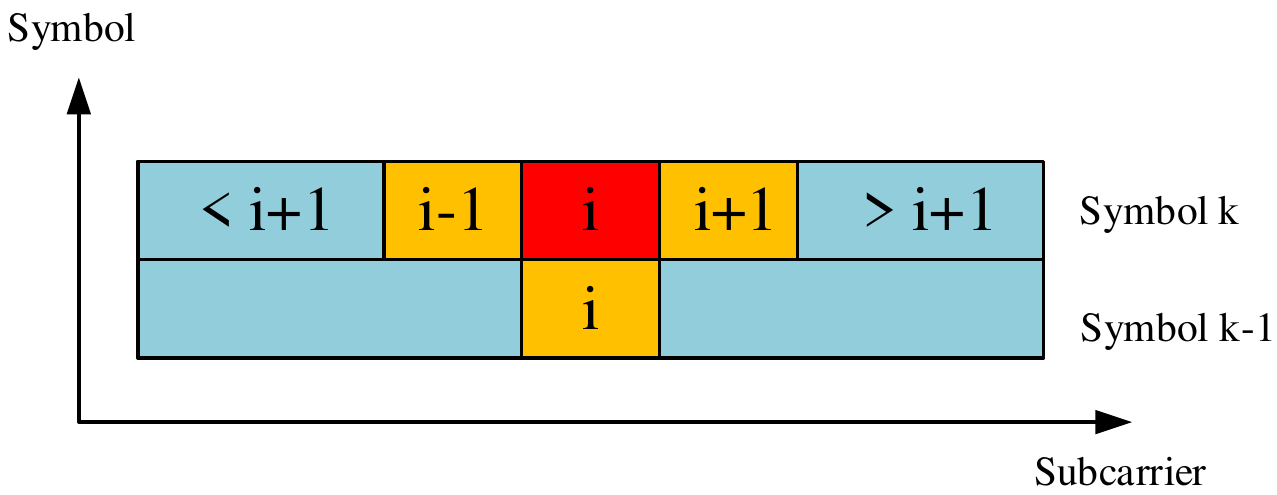}\vspace*{-0mm}
		\caption{Subcarrier $i$ of symbol $k$ is mainly interfere by the ICI and ISI originated from subcarriers $\mathcal{S}_\mathrm{c} = \{i-1,i+1\}$ of symbol $k$ and subcarrier $\mathcal{S}_\mathrm{p} = \{i\}$ of symbol $k-1$. The span of $\mathcal{S}_\mathrm{c}$ and $\mathcal{S}_\mathrm{p}$ can be extended at the cost of higher computational complexity.}
		\label{fig:interference}\vspace*{-0mm}
	\end{figure}
	
	Nevertheless, it is reported in \cite{Cisek18Frequency} that, for intended subcarrier $i$ in symbol $k$, the major power of ICI comes from the neighbouring subcarrier set $\mathcal{S}_\mathrm{c}$ of its own symbol. The ISI is mainly caused by subcarriers adjacent to subcarrier $i$ of the previous symbol, and the ISI subcarrier set is denoted as  $\mathcal{S}_\mathrm{p}$. This is illustrated in Figure \ref{fig:interference}. Therefore, we first manage to cancel the interference from subcarriers out of $\mathcal{S}_\mathrm{c}$ and $\mathcal{S}_\mathrm{p}$, and then perform joint ML estimation over subcarriers belongs to $\mathcal{S}_\mathrm{c} \bigcup \mathcal{S}_\mathrm{p}$. In particular, we rewrite the channel matrix $\mathbf{A}_k$ in \eqref{Ak} as:
	\begin{eqnarray}
	\mathbf{A}_k = [\bm{\beta}_{k-1,1}, \ldots, \bm{\beta}_{k-1,N}, \bm{\beta}_{k,1}, \ldots, \bm{\beta}_{k,N}],
	\end{eqnarray}
	where the vector $\bm{\beta}_{k,j} \in \mathbb{C}^{2N N_{\mathrm{R}}\time 1}, j\in\{1,\ldots,N N_{\mathrm{T}}\}$, denotes the column corresponding to symbol $k$ on subcarrier $i$ of matrix $\mathbf{A}_k$. Then, for estimating symbol $k$ on subcarrier $i$, we first cancel the interference caused by subcarriers out of $\mathcal{S}_\mathrm{c} \bigcup \mathcal{S}_\mathrm{p}$ from the received signal $\mathbf{R}_k$, and the resulting signal can be expressed as:
	\begin{eqnarray}
	\hspace*{-6mm} \widetilde{\mathbf{R}}_{k,i} \hspace*{-2mm} &=& \hspace*{-2mm} \mathbf{R}_k \notag \\
	\hspace*{-2mm} &-& \hspace*{-2mm} \underbrace{[\ldots,\bm{\beta}_{k-1,i\notin \mathcal{S}_\mathrm{p}}, \ldots, \bm{\beta}_{k-1,i\notin \mathcal{S}_\mathrm{c}},\ldots]
		\begin{bmatrix}
		\vdots\\
		\overline{X}_{k-1,i\notin \mathcal{S}_\mathrm{p}}\\
		\vdots \\
		\overline{X}_{k,i\notin \mathcal{S}_\mathrm{c}}\\
		\vdots
		\end{bmatrix},}_{\mbox{ICI and ISI caused by subcarriers out of } \mathcal{S}_\mathrm{c} \bigcup \mathcal{S}_\mathrm{p}}
	\end{eqnarray}
	where $\overline{X}_{k-1,i\notin \mathcal{S}_\mathrm{p}}$ denotes the decision constellation after pre-detection. 
	Then, we define the following matrices:
	\begin{eqnarray}
	\mathbf{A}_{k,i}^{\mathrm{cut}} & = & [\bm{\beta}_{k-1,i\in \mathcal{S}_\mathrm{p}}, \bm{\beta}_{k,i\in \mathcal{S}_\mathrm{c}}],\\
	\mathbf{Z}_{k,i}^{\mathrm{cut}} & = &
	\begin{bmatrix}
	X_{k-1,i\in \mathcal{S}_\mathrm{p}}\\
	X_{k,i\in \mathcal{S}_\mathrm{c}}
	\end{bmatrix}.
	\end{eqnarray}
	Now, the signal model after partial ICI and ISI cancellation is given by:
	\begin{eqnarray} \label{cutmodel}
	\widetilde{\mathbf{R}}_{k,i} = \mathbf{A}_{k,i}^{\mathrm{cut}} \mathbf{Z}_{k,i}^{\mathrm{cut}}.
	\end{eqnarray}
	As a result, the ML estimation for signal model \eqref{cutmodel} can be obtained by solving the following problem:
	\begin{eqnarray} \label{cutML}
	\hat{\mathbf{Z}}_{k,i} = \underset{\mathbf{Z}_{k,i} \in \mathcal{Z}_{\mathrm{QAM}}}{\arg \min} \,\, \norm{\widetilde{\mathbf{R}}_{k,i} - \mathbf{A}_{k,i}^{\mathrm{cut}} \mathbf{Z}_{k,i}^{\mathrm{cut}}}^2.
	\end{eqnarray}
	We note that \eqref{cutML} presents a joint ML detection for symbol $k$ in subcarrier $\mathcal{S}_\mathrm{c} \bigcup \mathcal{S}_\mathrm{p}$, which entails a smaller searching space than the original ML detection problem in \eqref{origML}. However, finding the optimal solution of \eqref{cutML} is still difficult due to the discrete feasible set. Thus, we exploit deep learning-based approach for solving \eqref{origML}. In particular, we utilize deep-unfolding method where each layer of the DNN mimics one iteration of the gradient descent method for solving \eqref{origML}. 
	Specifically, the output of layer $l$, i.e. $\hat{\mathbf{Z}}_{k,i}^{(l+1)}$, is expected to be modeled as the function of the output of the previous layer and the first order partial derivatives of the objective function of \eqref{cutML}, which is given by:
	\begin{eqnarray}
	&& \hspace*{-5mm} \hat{\mathbf{Z}}_{k,i}^{(l+1)}  \\
	& = & \Pi \Bigg[\hat{\mathbf{Z}}_{k,i} ^{(l)} - \delta_k \frac{\partial {\norm{\widetilde{\mathbf{R}}_{k,i} - \mathbf{A}_{k,i}^{\mathrm{cut}} \mathbf{Z}_{k,i}^{\mathrm{cut}}}}^2}{\partial \mathbf{Z}_{k,i}^{\mathrm{cut}}} \arrowvert_{\mathbf{Z}_{k,i}^{\mathrm{cut}} = \hat{\mathbf{Z}}_{k,i}^{(l)}}\Bigg] \notag \\
	&=& \Pi\Big[\hat{\mathbf{Z}}_{k,i} ^{(l)} - \delta_k {\mathbf{A}_{k,i}^{\mathrm{cut}}}^T \widetilde{\mathbf{R}}_{k,i} + \delta_k {\mathbf{A}_{k,i}^{\mathrm{cut}}}^T \mathbf{A}_{k,i}^{\mathrm{cut}}  \hat{\mathbf{Z}}_{k,i}^{(l)} \Big], \notag
	\end{eqnarray}
	where $\Pi[x]$ and $\delta_k$ denotes the mapping function and step size, respectively.
	In practice, we expect to train the DNN to approximate $\Pi[x]$ and $\delta_k$, which can be expressed as:
	\begin{eqnarray}
	\hat{\mathbf{Z}}_{k,i}^{(l+1)} = \psi_t\Bigg(\mathbf{W}_l 
	\begin{bmatrix}
	{\mathbf{A}_{k,i}^{\mathrm{cut}}}^T \widetilde{\mathbf{R}}_{k,i} \\
	{\mathbf{Z}_{k,i}^{\mathrm{cut}}}^{(l)}\\
	{\mathbf{A}_{k,i}^{\mathrm{cut}}}^T \mathbf{A}_{k,i}^{\mathrm{cut}}
	\end{bmatrix}
	+ \mathbf{b}_l
	\Bigg),
	\end{eqnarray}
	where $\mathbf{W}_l$ and $\mathbf{b}_l$ denotes the trainable weights and bias of the DNN, respectively.
	$\psi_t(x)$ denotes the active function which is defined as:
	\begin{eqnarray}
	\psi_t(x) = P\Big[-1 + \frac{\mathrm{Relu}(x+t)}{\abs{t}} - \frac{\mathrm{Relu}(x-t)}{\abs{t}} \Big],
	\end{eqnarray}
	where $\mathrm{Relu}(x) = \max(0,x)$, $t$ is the trainable function, $P$ denotes the maximum absolute value of the power normalized M-QAM constellation sets. The loss function of the proposed DNN is defined as
	\begin{eqnarray}
	\mathrm{Loss} = \sum_{l=1}^{L} \log(l) {\norm{\mathbf{Z}_{k,i} - \hat{\mathbf{Z}}_{k,i}^{(l+1)}}}^2.
	\end{eqnarray}

	\begin{figure}
		\centering\vspace*{-2mm}
		\includegraphics[width=3.3in]{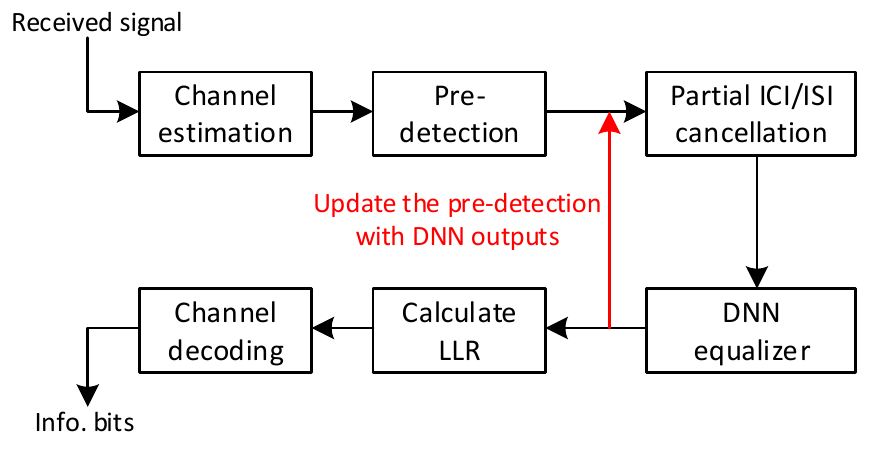}\vspace*{-2mm}
		\caption{Iterative receiver structure.}
		\label{fig:receiver}\vspace*{-3mm}
	\end{figure}

	We note that the performance of the proposed DNN equalization is impact by the accuracy of the pre-detection $\overline{X}_{k,i}$. Thus, we can construct a iterative receiver which  update the pre-detection symbols with the output of the DNN after the detecting of the symbols in a frame. The iterative receiver structure is shown in Figure \ref{fig:receiver}.

	\section{Performance Evaluation}

	\begin{table}[t]\vspace*{-0mm}\caption{System parameters}\vspace*{-2mm}\label{tab:parameters} 
		\newcommand{\tabincell}[2]{\begin{tabular}{@{}#1@{}}#2\end{tabular}}
		\centering
		\begin{tabular}{|l|l|}\hline
			\hspace*{-1mm}Carrier center frequency  & $4$ GHz  \\
			\hline
			\hspace*{-1mm}Subcarrier spacing & $30$ kHz \\
			\hline
			\hspace*{-1mm}Number of resource block  (RB) & $4$  \\
			\hline
			\hspace*{-1mm}Number of subcarriers per RB & $12$ \\
			\hline
			\hspace*{-1mm}Number of FFT points & $2048$\\
			\hline
			\hspace*{-1mm}The length and number of sampling point of CP & $2.3$ us and $144$\\
			\hline
			\hspace*{-1mm}Number of symbols per frame   &  $14$  \\
			\hline
			\hspace*{-1mm}Number of users  &  $2$   \\
			\hline
			\hspace*{-1mm}Number of antennas and data streams at each user  &  $2$ and $2$  \\
			\hline
			\hspace*{-1mm}User moving speed   &   \mbox{$3$ km/h}  \\
			\hline
			\hspace*{-1mm}Number of layers of the proposed neural network   &   $30$  \\
			\hline
		\end{tabular}
		\vspace*{-3mm}
	\end{table}

	\begin{table}[t]\vspace*{-0mm}\caption{Measured Power Delay Profile}\vspace*{-2mm}\label{tab:PDP} 
		\newcommand{\tabincell}[2]{\begin{tabular}{@{}#1@{}}#2\end{tabular}}
		\centering
		\begin{tabular}{l|c|c|c|c|c|c|c}\hline
			Path index & 1 & 2 & 3 & 4 & 5 & 6 & 7 \\
			\hline
			Delays (us) & 0 & 0.135 & 0.534 & 0.681 & 2.404 & 3.766 & 4.553\\
			\hline
			Avg. gain (dB) & 0 & 0.790 & 3.531 & 3.123 & 0.456 & 3.700 & 0.474\\
			\hline
		\end{tabular}
		\vspace*{-3mm}
	\end{table}

	In this section, we investigate the detection performance of the proposed receiver design through simulations. The adopted the simulation parameters and DNN training parameters are given in Table \ref{tab:parameters}, unless specified otherwise.
	In this paper, we consider two types of channel models. The first is long multipath channels whose delay spread exceeding the length of CP. The long multipath channel is modeled according to the measured power delay profile given in Table \ref{tab:PDP} and the fading coefficients of each path are generated as independent and identically distributed Rayleigh random variables. The second type of channels are modeled according to the tapped delay line-A (TDL-A) channel model  \cite{cho2010mimo} and their maximum path delay is shorter than the length of CP.
	The DNN is initialized by setting $\hat{\mathbf{Z}}_{k,i}^{(l+1)} = \mathbf{0}$.
	ADMM optimizer is used for training DNN with learning rate $0.0001$. The simulation was performed on the 5G NR link-level platform where LDPC is considered for channel coding. The DNN is trained in a supervised manner over 10 thousand channel realizations.
	
	For comparison, we consider two baseline equalization schemes. The first is the traditional MMSE equalization, where the the received signal on each subcarrier is equalized separately. The second baseline is the modified ZF equalization derived from the signal model in \eqref{Yk}. In particular, all subcarrier are equalized jointly by $\mathbf{V}_k^{\mathrm{ZF}} = \mathbf{D}_k^H(\mathbf{D}_k \mathbf{D}_k^H)^{-1}$, where $\mathbf{D}_k = \mathbf{H}_k - \bm{\Phi}_k \mathbf{B}$.
	
	\begin{figure}
		\centering\vspace*{-0mm}
		\includegraphics[width=3.3in]{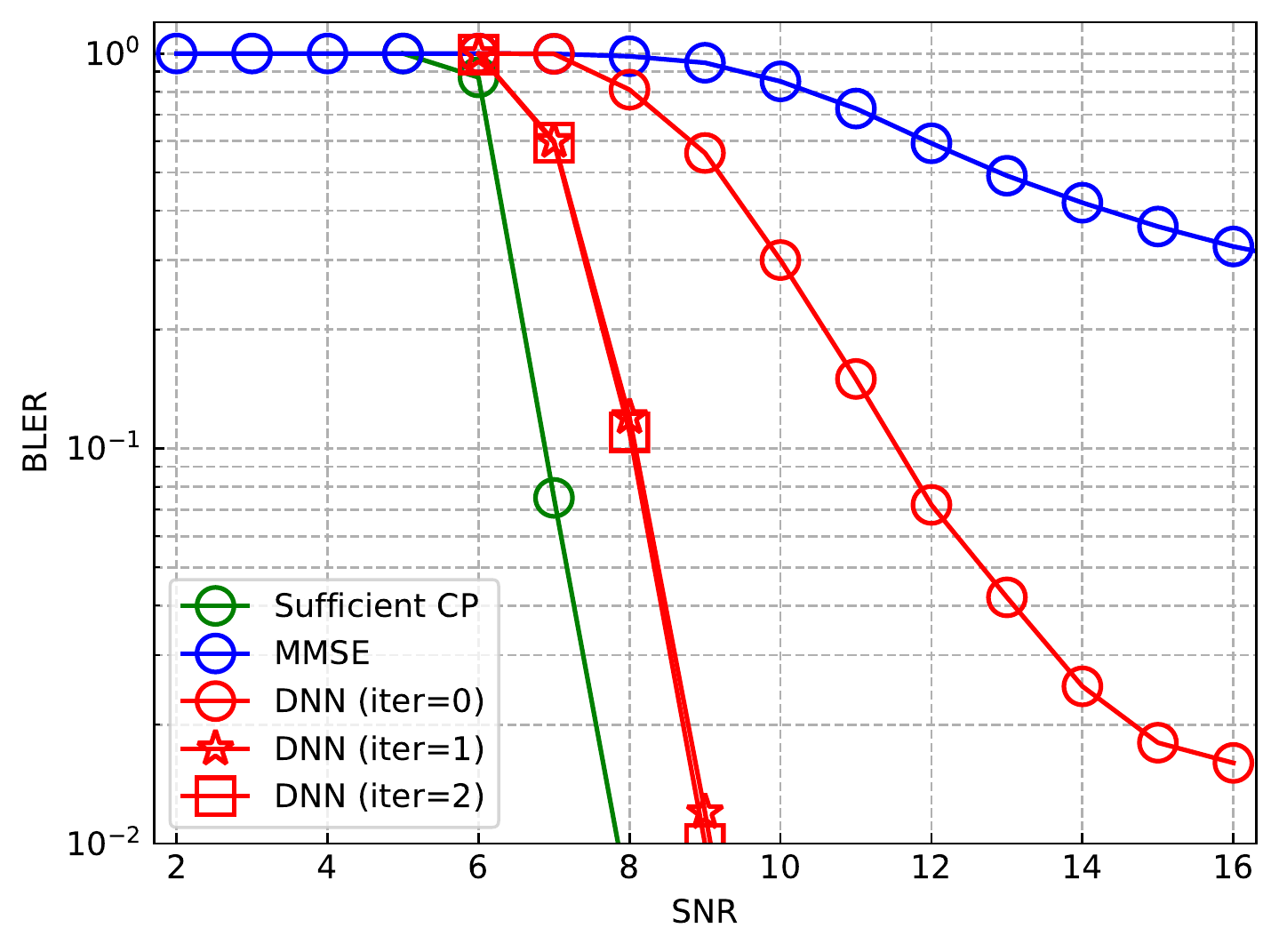}\vspace*{-2mm}
		\caption{Average BLER versus SNR for two long-path-delay users with MCS19 and perfect CSI.}
		\label{fig:AllFinland_MCS19_ICE}\vspace*{-3mm}
	\end{figure}
	Figure \ref{fig:AllFinland_MCS19_ICE} illustrates the average block error rate (BLER) versus SNR, for MCS19 (64QAM) with two long-path-delay users and perfect channel state information (CSI). Normally, we focus on the minimum required SNR value for achieving target BLER $0.1$. As can be observed, the traditional MMSE scheme cannot work in such a harsh scenario while the proposed DNN equalizer achieves a great BLER improvement than MMSE. The detection performance can be further improved in one more detection iteration with the proposed receiver. The curve marked by \emph{Sufficient CP} is obtained by prolonging the length of CP artificially using traditional MMSE equalization and serves as the upper bound of the performance.

	\begin{figure}
		\centering\vspace*{-0mm}
		\includegraphics[width=3.45in]{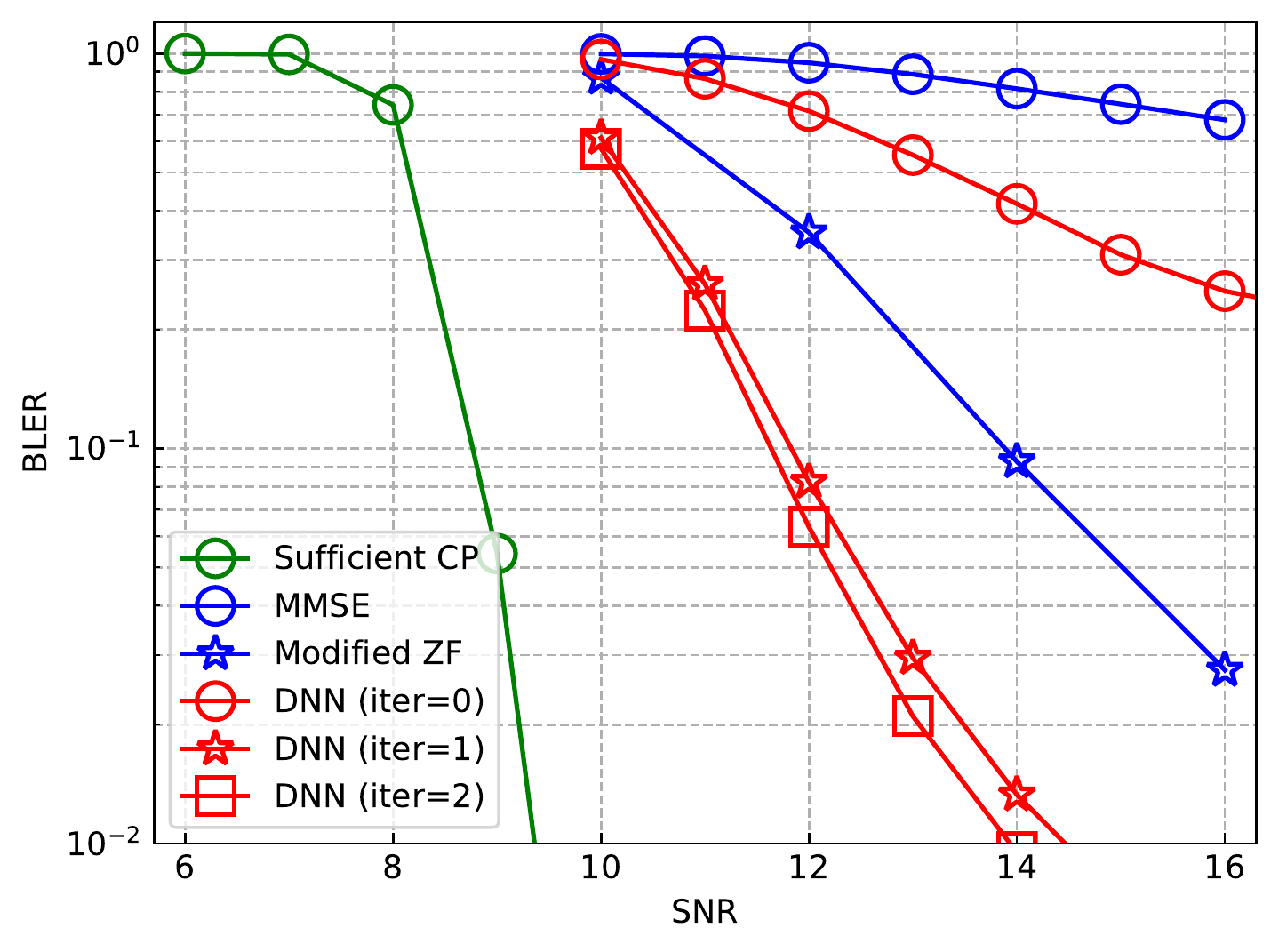}\vspace*{-2mm}
		\caption{Average BLER versus SNR for two long-path-delay users with MCS19 and imperfect CSI.}
		\label{fig:AllFinland_MCS19_RCE}\vspace*{-3mm}
	\end{figure}
	In Figure \ref{fig:AllFinland_MCS19_RCE}, we investigate the detection performance with imperfect CSI obtained via least-square (LS) channel estimation scheme. As can be observed, after performing two iterations, the proposed receiver can achieve significant performance improvement compared to MMSE, and $2$ dB gain compared to the modified ZF. We note that the modified ZF is a favourable scheme which entails high computational complexity.

	\begin{figure}
		\centering\vspace*{-0mm}
		\includegraphics[width=3.45in]{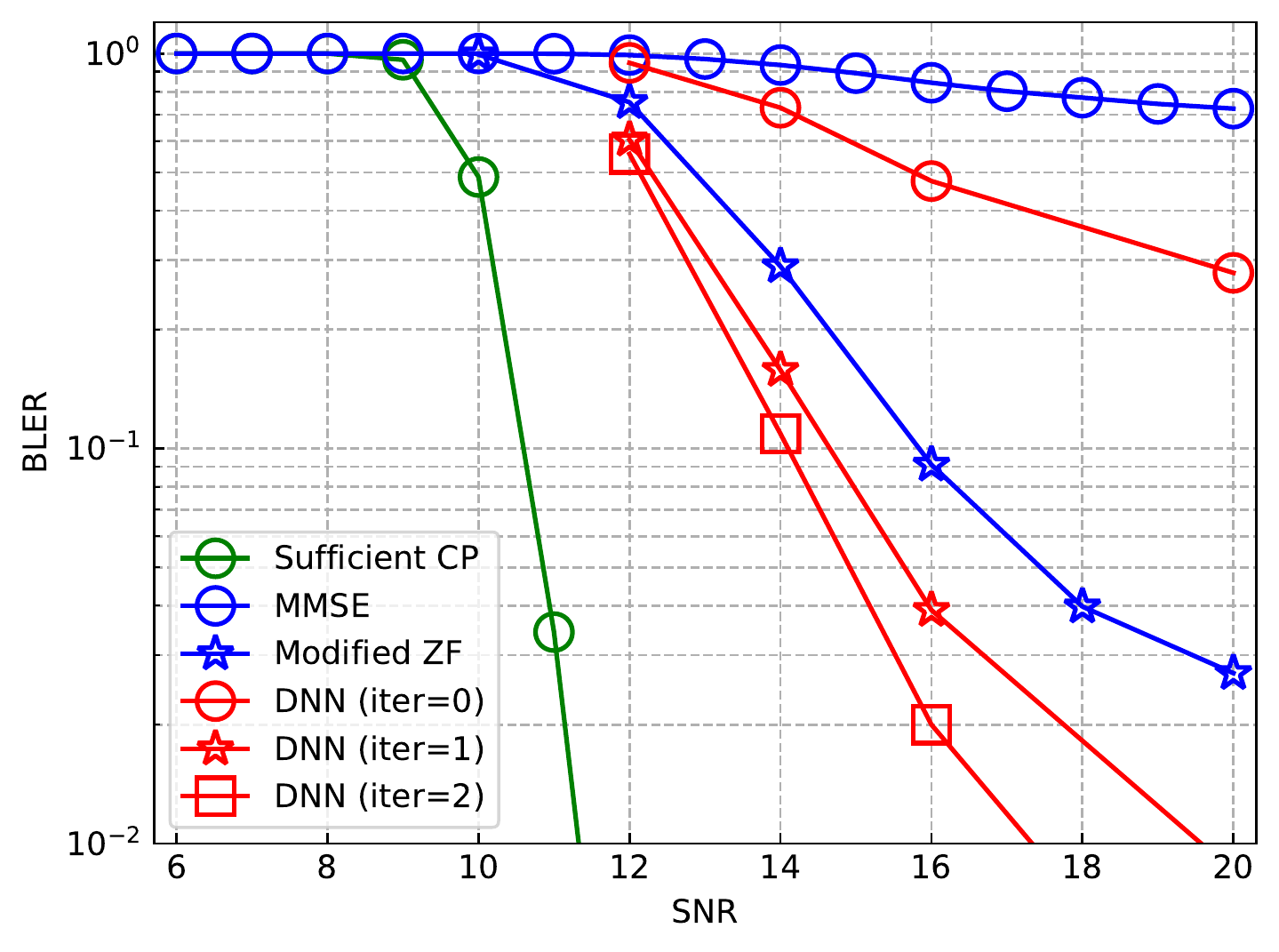}\vspace*{-2mm}
		\caption{Average BLER versus SNR for the long-path-delay user with MCS20 and imperfect CSI.}
		\label{fig:FinTDLA_MCS20_RCE_FinUE}\vspace*{-3mm}
	\end{figure}
	
	\begin{figure}
		\centering\vspace*{-0mm}
		\includegraphics[width=3.45in]{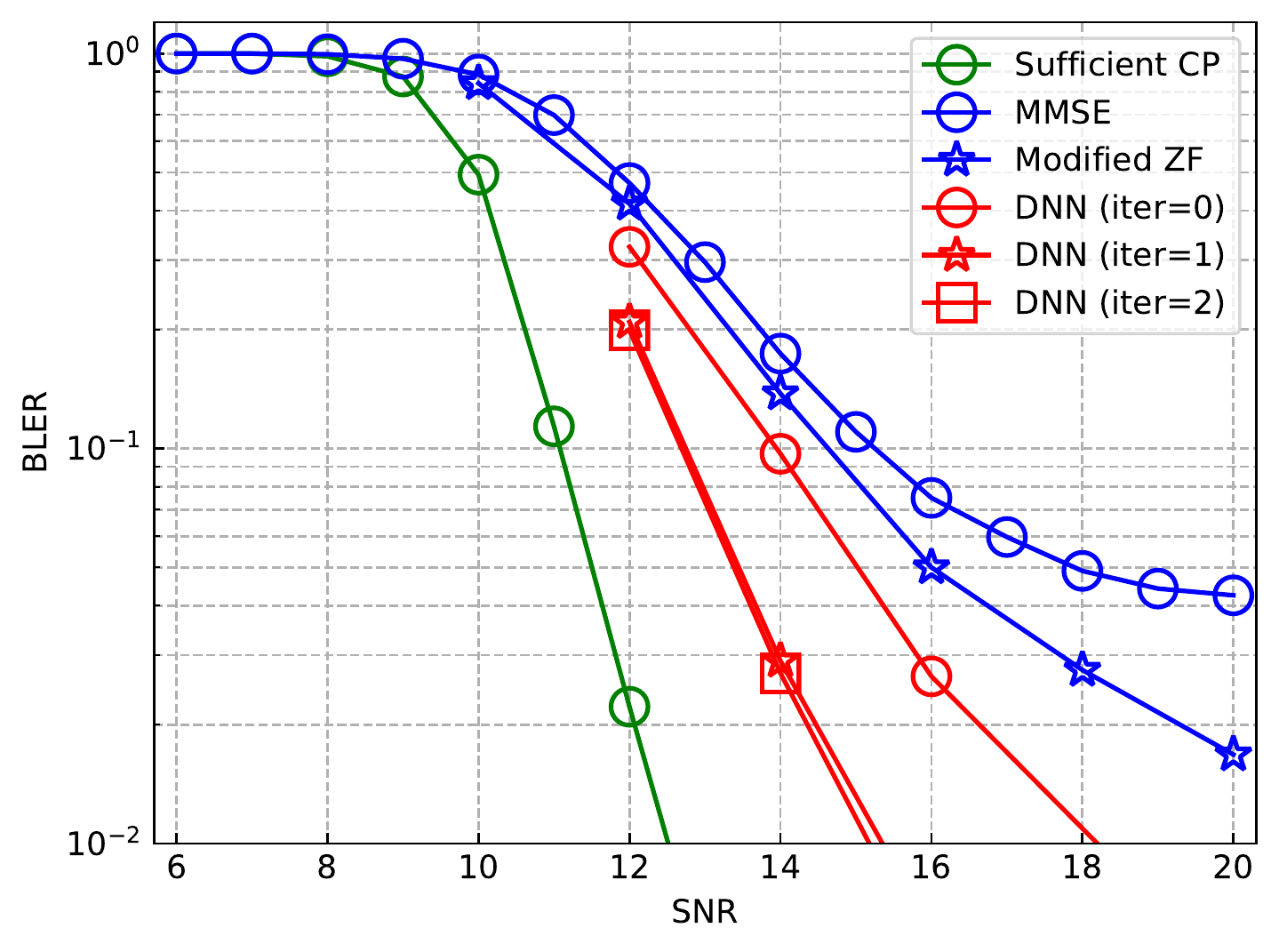}\vspace*{-2mm}
		\caption{Average BLER versus SNR for the TDLA user with MCS20 and imperfect CSI.}
		\label{fig:FinTDLA_MCS20_RCE_TDLAUE}\vspace*{-3mm}
	\end{figure}
	In Figure \ref{fig:FinTDLA_MCS20_RCE_FinUE} and Figure \ref{fig:FinTDLA_MCS20_RCE_TDLAUE}, we evaluate the detection performance with higher constellation order, i.e., MCS20 (256QAM), with one long-path-delay user and one TDLA user for imperfect CSI. In particular, Figure \ref{fig:FinTDLA_MCS20_RCE_FinUE} illustrate the performance of the long-path-delay user, and the proposed DNN receiver outperforms the traditional MMSE and modified ZF as expected. Interestingly, we also note that the proposed receiver can 
	improve the performance of the TDLA user who has sufficient CP. 
	
	
	\vspace*{-2mm}
	\section{Conclusions}
	In this paper, we studied the deep learning based equalization design for MIMO-OFDM systems where the multipath delay spread exceeding the length of CP. The proposed equalizer was designed to solve an ML detection problem via unfolding approach. Inspired by the observation that a given subcarrier is mainly interfered by its adjacent subcarriers, a low-complexity scheme was proposed by cancelling the interfering subcarrier that far away from the intended one. An iterative receiver design was also proposed for 
	reducing error of interference cancellation. Simulation results confirmed that the detection performance is greatly enhanced even with one iteration.
	In addition, our results revealed that a substantial improvement of detection gain can be achieved by employing the proposed equalization scheme compared to conventional detection schemes.
	
	\vspace*{-2mm}

\end{document}